\documentclass[review]{elsarticle}

\usepackage{lineno,hyperref,amssymb,graphicx,booktabs}
\modulolinenumbers[5]

\journal{Nuclear Instruments and Methods in Physics Research A}

\begin{document}

\begin{frontmatter}

\title{Large Magnetic Shielding Factor Measured by Nonlinear
  Magneto-optical Rotation}

\author[winnipeg,manitoba]{J.W. Martin\corref{mycorrespondingauthor}}
\cortext[mycorrespondingauthor]{Corresponding author}
\ead{j.martin@uwinnipeg.ca}
\author[winnipeg,manitoba]{R.R. Mammei}
\author[winnipeg,manitoba]{W. Klassen}
\author[winnipeg]{C. Cerasani}
\author[manitoba]{T. Andalib}
\author[winnipeg,manitoba]{C.P. Bidinosti}
\author[manitoba]{M. Lang}
\author[winnipeg]{D. Ostapchuk}

\address[winnipeg]{Physics Department, The University of Winnipeg, 515 Portage Avenue, Winnipeg, MB, R3B 2E9, Canada}
\address[manitoba]{Department of Physics and Astronomy, University of Manitoba, Winnipeg, MB R3T 2N2, Canada}

\begin{abstract}
A passive magnetic shield was designed and constructed for
magnetometer tests for the future neutron electric dipole moment
experiment at TRIUMF.  The axial shielding factor of the magnetic
shield was measured using a magnetometer based on non-linear
magneto-optical rotation of the plane of polarized laser light upon
passage through a paraffin-coated vapour cell containing natural Rb at
room temperature.  The laser was tuned to the Rb D1 line, near the
$^{85}$Rb $F=2\rightarrow 2,3$ transition.  The shielding factor was
measured by applying an axial field externally and measuring the
magnetic field internally using the magnetometer.  The axial shielding
factor was determined to be $(1.3\pm 0.1)\times 10^{7}$, from an
applied axial field of 1.45~$\mu$T in the background of Earth's
magnetic field.
\end{abstract}


\end{frontmatter}


\section{Introduction}

The next generation of neutron electric dipole moment (EDM)
experiments aim to measure the EDM $d_n$ with proposed precision
$\delta d_n\lesssim
10^{-27}$~e-cm~\cite{bib:nedm1,bib:nedm2,bib:nedm2.5,bib:nedm3,bib:nedm3.5,bib:nedm4,bib:nedm5,bib:nedm6,bib:nedm6.5}.
In the previous best experiment \cite{bib:baker}, which discovered
$d_n<2.9\times 10^{-26}$~e-cm, effects related to magnetic field
homogeneity and instability were found to dominate the systematic
error.  A detailed understanding of passive and active magnetic
shielding, magnetic field generation within shielded volumes, and
precision magnetometry is expected to be crucial to achieve the
systematic error goals for the next generation of experiments.  Much
of the R\&D effort for these experiments is focused on careful design
and testing of various magnetic shield geometries with precision
magnetometers~\cite{bib:brys,bib:afach,bib:fierlingerroom,bib:sturmthesis,bib:patton}.

Our work focused on the creation of a new small-scale magnetic shield
designed primarily for magnetometer tests for our future nEDM
experiment at TRIUMF.  Within the shield, we prepared a magnetometer
based on non-linear magneto-optical rotation (NMOR).  NMOR results in
a rotation of the plane of polarization of laser light resonant with
atomic transitions in Rb vapour.  In NMOR, the optical properties of
the medium are modified by the laser light, resulting in nonlinear
effects such as hole-burning and the creation of a coherent dark
state~\cite{bib:budkeramj}.  The combined effect results in
ultranarrow linewidths of $\sim$Hz, corresponding to projected field
sensitivities at the few fT level~\cite{bib:budker1998}.  The effect,
which occurs near zero field, can be displaced to a non-zero field
using either a frequency-modulated (FM) \cite{bib:budkerfm} or
amplitude-modulated (AM) \cite{bib:budkeram,bib:higbie,bib:hovde} pump
beam, and a CW probe beam, and can in principle retain $\sim$fT
precision~\cite{bib:jacksonkimballfm}.  At such precision, the
technology is superior to fluxgate magnetometers, and rivals the
precision of SQUID magnetometers, without the need for cryogenics.
Recently, a three-axis NMOR-based magnetometer has been
demonstrated~\cite{bib:patton}.

Our magnetometer was based on the zero-field effect along a single
axis.  By calibrating the zero-field magnetometer using coils internal
to the magnetic shield, we determined the axial magnetic shielding
factor for fields applied by external coils.  A key result of this
work is that very large DC shielding factors could be measured in a
small space, for small applied external fields (more than an order of
magnitude smaller than Earth's field).  Such measurements could not
have been conducted using a conventional fluxgate magnetometer.  The
results validate our initial design goals for the magnetic shielding.
The results also highlight the applicability of precision NMOR-based
magnetometers in magnetically shielded environments, such as those
that will be seen in future neutron EDM experiments.

\section{Passive Magnetic Shield System}
Analytical approximations exist for the transverse and axial shielding
factors of completely closed and open finite multi-layered cylindrical
shapes made of a high permeability $\mu$
material~\cite{bib:gubser1979,bib:dubbers1986,bib:sumner1987,bib:paperno2000}.
Here transverse (axial) shielding implies a reduction in the
externally applied magnetic field that is perpendicular (parallel) to
the axis of concentric cylindrical shells.  Closed-ended cylindrical
structures have a higher axial shielding factor than open-ended
concentric shells~\cite{bib:gubser1979}.  Shield structures often need
apertures permitting access to the internal shielded volume for
e.g. laser light, subatomic particles, wires for field generation,
etc.  Often cylindrical extensions, or ``stovepipes'', are situated
around the apertures.  Stovepipes generally may increase the shielding
factor, however the optimum length is highly dependent on the design
constraints and parameters of the shield structure.  The shield used
in this work was designed using a commercial finite element analysis
(FEA) software package~\cite{bib:opera}.  Special care was taken to
design the apertures and stovepipes in the ends of the shield
achieving as large as possible axial magnetic shielding factor.

\subsection{Design Constraints}

The shield was designed in using FEA by requiring that the magnetic
field within a region of interest (ROI) internal to the shield be as
small as possible under application of an external axial field.  For
design purposes, the ROI was selected to be 4'' (10.16~cm) long and
1'' (2.54~cm) diameter, corresponding to the space inhabited by most
magnetometers of interest.  The innermost shield was required to be at
least 4'' (10.16 cm) in diameter and 8'' (20.32~cm) long, in order to
accommodate various coils as well as the magnetometers and support
structures.  The outermost shield diameter was required to be less
than 12'' (30.48~cm) to keep the shield small enough to rest on an
optics table.  All access to the inner shield was to be provided by a
single $1\frac{1}{8}$'' (2.86~cm) protrusion on each end to admit
laser light along the axis (as well as cabling for field coils).  Only
the axial shielding factor was considered in designing the shield,
because the axial shielding factor is generally smaller than the
transverse shielding factor for this geometry.

The material to be used for the shield was fixed at 1/16" (0.159~cm)
thick, since this material is readily available and cost effective.
Both three- and four-layer shields were considered.  Initial estimates
of the axial shielding factor were conducted by comparing the results
of the approximate axial shielding factors of
Refs.~\cite{bib:gubser1979}.  A stovepipe geometry on the axial
2.86~cm protrusion in each shield layer was considered based on the
analytical result found in Ref.~\cite{bib:Mager1967} for the
exponential decay of an axial field leaking into an open-ended
cylinder.

A starting design based on these considerations was then implemented
in the FEA simulation.  The length $L_i$ and radius $R_i$ of each
shielding layer $i$, the number of layers ($N=3$ or $4$ where $1\leq
i\leq N$ and $i=1$ refers to the innermost shield), and the length of
the stovepipe $\ell$ were allowed to vary in the simulation.  The
radii of the shields were constrained to obey $R_{i+1}/R_i=c$ where
$c$ is a constant independent of the layer.  The length of each
successive layer was constrained to follow
$L_{i+1}=L_{i}+2\ell+R_{i+1}-R_{i}$.  Thus there is a gap between the
end of a stovepipe and the next layer's lid of $\frac{1}{2}(R_{i+1}-
R_{i})$.  This ensures sufficient space between the stovepipes and
lids of each layer, i.e. the stovepipes are not nested.  Other
fractions of this gap were not considered.  As expected based on
simple estimation, the best results for the shielding factor were
found for the larger number of shield layers $N=4$ and the value of
$c$ being as large as possible $c=(R_4/R_1)^{1/3}=1.4$ so that
$R_4=6$'' if $R_1$ is constrained to be 2''.

\begin{figure}
\begin{center}
\includegraphics[width=4in]{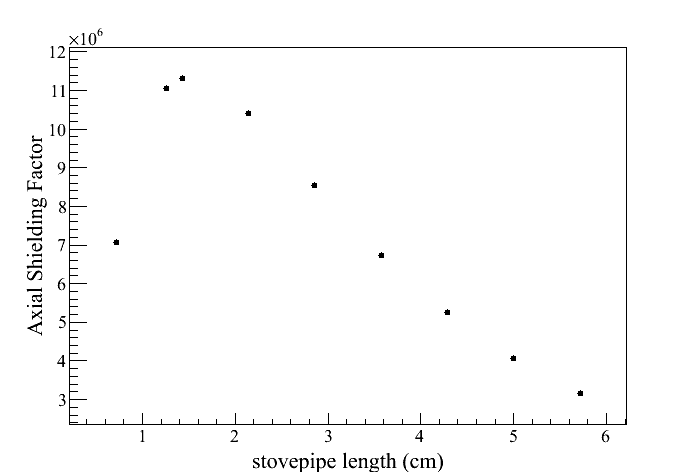}
\caption{FEA study of axial shielding factor vs. stovepipe length
  $\ell$ for a four-layer shield with radial dimension as in
  Fig.~\ref{fig:design}.  Each stovepipe of the next shield was
  required to be fully contained within the next shield.  Under these
  conditions, the axial shielding factor decreases as the stovepipe
  length (and therefore overall shield length) increases.  The optimal
  stovepipe length is of order half the hole
  diameter.\label{fig:stovepipevariation}}
\end{center}
\end{figure}

A somewhat surprising result was that shorter shields with shorter
stovepipes were preferable, in that the additional length implied by
longer stovepipes ultimately served only to make the shield longer,
with a somewhat negative impact on the overall shielding factor.  The
FEA result is displayed graphically in
Fig.~\ref{fig:stovepipevariation}.  This was found to be qualitatively
in agreement with recent work in Ref.~\cite{bib:burt2002}.

Magnetic field homogeneity using a simple solenoidal internal coil
design (reminiscent of the one described in Section~\ref{sec:coils})
was also checked in the ROI, and confirmed that the 20.32 cm length of
the innermost shield would be sufficient for sub-percent homogeneity
over the ROI.

\subsection{Final Design and Fabrication}

\begin{figure}
\begin{center}
\includegraphics[width=4in]{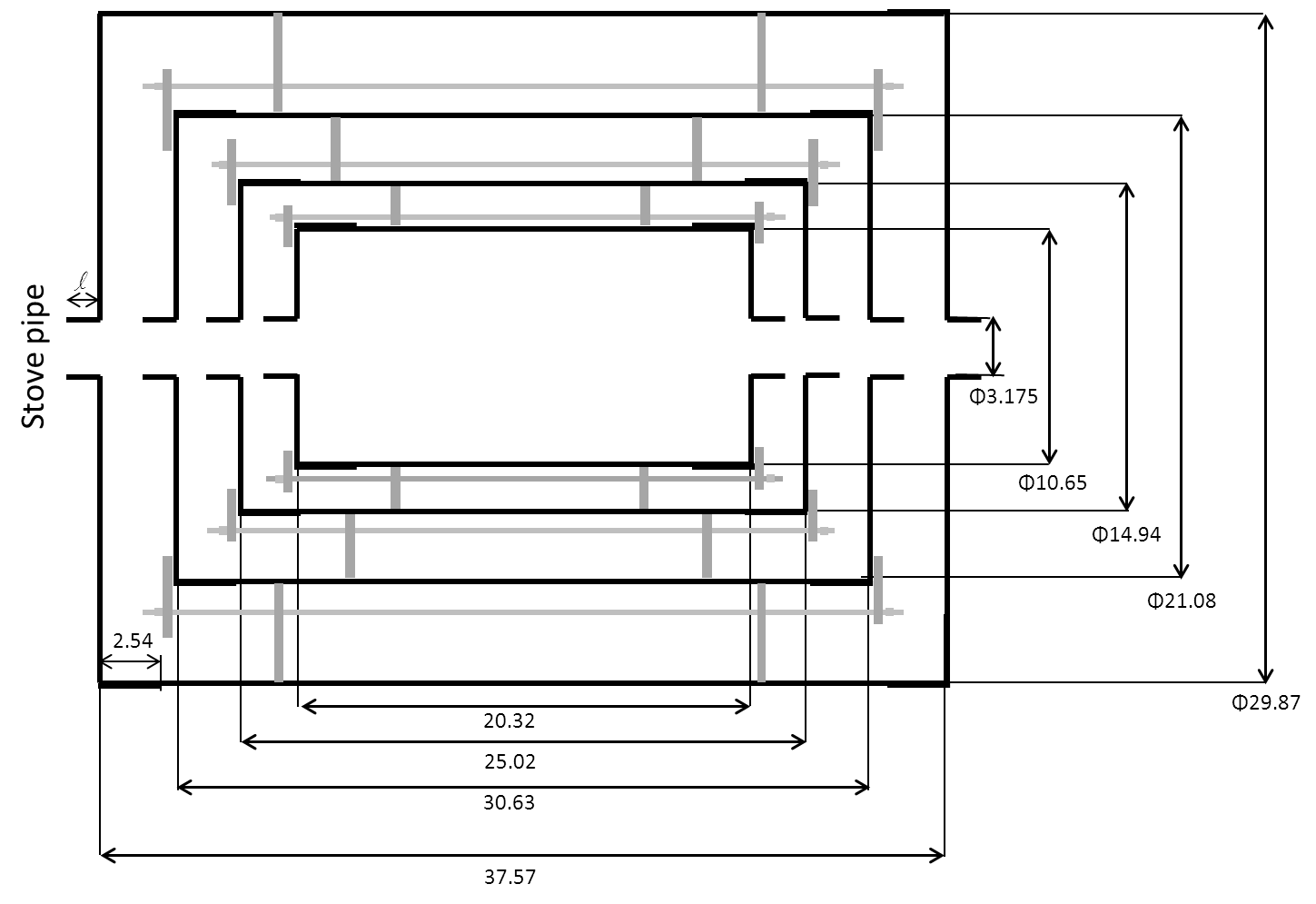}
\caption{Schematic diagram of the 4-layer magnetic shield (dimensions
  in cm).  Outer diameters of the cylinders are shown, and dimensions
  are quoted in centimeters.  The diameter of each endcap is larger by
  0.1~cm to fit over its corresponding cylinder, extending 2.54~cm
  over the end of it.  The hole diameter and stovepipe length for each
  endcap are the same.  High density polyethylene spacers and nylon
  thread rods/nuts, shown in gray, hold the shields and end caps
  together.\label{fig:design}}
\end{center}
\end{figure}

The final design is shown in Fig.~\ref{fig:design} and the dimensions
as constructed were similar.  The magnetic shield was fabricated and
annealed by Ad-Vance Magnetics, Inc.~\cite{bib:ad-vance}.  The vendor
informally suggested an effective DC permeability relative to that of
free space $\mu_r=20,000$ for the AD-MU-80 material used for the
shield.  Using this value in our FEA simulations implied an
anticipated axial magnetic shielding factor of $1.1\times 10^7$.

\subsection{Internal Coils \label{sec:coils}}

An internal coil, henceforth referred to as the $z$-coil, provided a
uniform field in the ROI, directed along the axis of the cylindrical
volume.  The $z$-coil was wound on a 7.62~cm diameter, 20.32~cm long
ABS plastic pipe.  Seven turns of 26~AWG magnet wire were wound at
2.54~cm spacing, with 1.27~cm spacing from the magnetic faces of the
endcaps of the innermost magnetic shield.  The spacings were chosen so
that, in the infinite permeability limit, and in the limit where the
axial aperture holes in the endcaps are small, the boundary conditions
would produce image currents forming an infinitely long
solenoid~\cite{bib:durand,bib:lambert}.  This is similar to the design
strategy used in Ref.~\cite{bib:budkeramj}.  Two saddle coils were
wound on the same cylinder in order to control transverse fields
internally; these were normally disconnected during precision
measurements.

The internal coil system was calibrated using a three-axis fluxgate
magnetometer at fields of $\sim$100~nT.  The calibration of the
$z$-coil was verified using the NMOR magnetometer with an AM pump
beam, and the known gyromagnetic ratios of Rb-85 and Rb-87 (described
in Section~\ref{sec:nmor} and similar to Ref.~\cite{bib:higbie}).

Homogeneity of the residual field and magnetic field generated by the
coil system was measured by scanning a fluxgate magnetometer along the
axis of the system with and without the coil energized.  At a field of
1~$\mu$T, the axial field generated by the coil was uniform within the
ROI to the 1\% level.

A single degaussing coil formed from a single loop of 12 AWG
multi-stranded insulated copper wire was fed through all four endcaps
and wound tight to both the innermost shield surface and outermost
shield surface.  In degaussing, a Variac was used to slowly ramp the
current in the degaussing coil, by hand, from zero to 10~A and down
again.  A 100~$\Omega$ power resistor in series with the degaussing
coil was used to limit the current the Variac delivered.  The system
provided zero magnetic field within the system reliably to the
$<0.3$~nT level.  Multiple degaussing trials would reduce this to the
$<0.1$~nT level, which was sufficient for the measurements conducted
using the NMOR-based magnetometer.  During precision magnetometer
measurements, the degaussing coil would be disconnected from its power
supply.

\section{NMOR magnetometer\label{sec:nmor}}

\begin{figure}
\begin{center}
\includegraphics[width=3in]{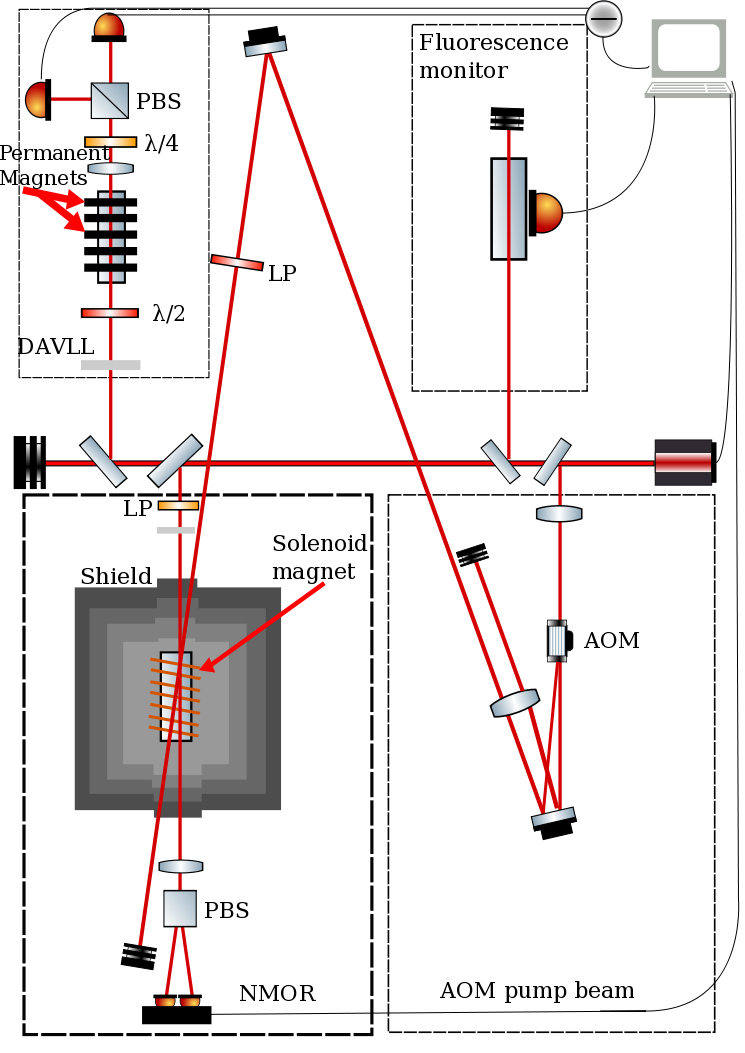}
\caption{Schematic diagram (top view) of magnetic shield and NMOR
  magnetometer system.  The main NMOR experiment is conducted by
  splitting off a fraction of the beam from the laser into the
  magnetic shield system and through the Rb vapour cell within.
  Optical rotation is measured using a polarizing beam splitter (PBS,
  being a Wollaston prism) followed by a balanced photosensor.  For
  the results in this publication, this beam served as both pump and
  probe.  An acousto-optic modulator (AOM) may be used to provide an
  amplitude-modulated pump beam, using the previously mentioned system
  as a probe, thereby creating a magnetometer capable of measuring
  fields far from $B=0$ (similar to Ref.~\cite{bib:higbie}); for this
  work, the pump beam was used only during a calibration
  cross-check on the magnetic field generated by the internal coil.
  The fluorescence monitor is used in beam tuning and the DAVLL system
  is used to lock the laser frequency.\label{fig:schematic}}
\end{center}
\end{figure}

A schematic diagram of the NMOR magnetometer system is shown in
Fig.~\ref{fig:schematic}.  Laser light was provided by a Toptica
DL-100 external cavity diode laser~\cite{bib:toptica}, tuned to the Rb
D1 line.  Light impinged upon a vapour cell containing natural Rb,
placed at the center of the coil and shield system.

The vapour cell was cylindrical, 5 cm long and 5 cm in diameter with
optical flats on the ends.  The cell was provided by D.~Budker, having
been prepared in a fashion similar to the cells described in
Ref.~\cite{bib:graf}.  The cell was characterized using a method
similar to Ref.~\cite{bib:graf}, by measuring the relaxation of
longitudinal polarization using optical rotation as a probe.  The long
time component of the relaxation was thereby found to decay with a
time scale of 0.36~s.

The temperature of the vapour cell was controlled by the ambient
temperature of the surrounding room ($\sim 21^\circ$C).  Transmitted
light was analyzed for optical rotation by a balanced polarimeter
system containing a Wollaston prism and a Newport model 2307 balanced
photoreceiver~\cite{bib:Newport}.

The power delivered to the vapour cell was typically 15 $\mu$W,
measured periodically using a Newport Model 818-SL power
meter~\cite{bib:Newport} inserted into the beamline.  The laser was
first tuned near the Rb-85 $F=2$ to $F'=2,3$ absorption minimum
(fluorescence maximum), then further tuned to maximize optical
rotation for $\sim 100$~pT applied fields.  The laser frequency was
stabilized using a dichroic atomic vapour laser lock (DAVLL) system
\cite{bib:wieman} containing optics for additional locking flexibility
based on Ref.~\cite{bib:yashchuk}, and using the Toptica DigiLock 110
system \cite{bib:toptica}.

\begin{figure}
\begin{center}
\includegraphics[width=\textwidth]{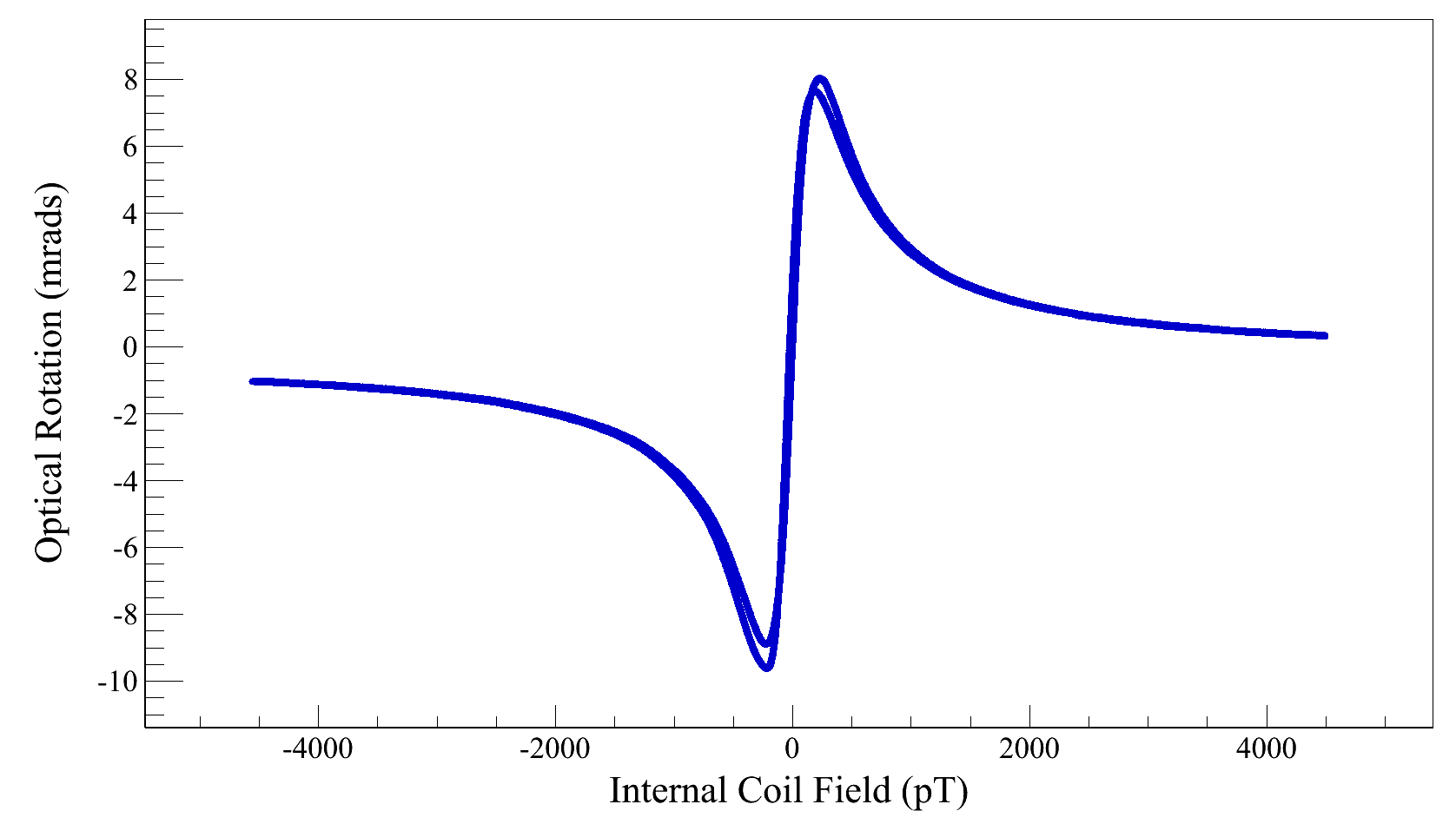}
\caption{Optical rotation as a function of magnetic field applied
  along the direction of the laser beam.  Data were acquired by
  sweeping the magnetic field in a triangle wave with a period of
  100~s; the peaks in the curve show a slight dependence on the
  direction of the sweep, due to the time response of the
  atoms.\label{fig:spectrum}}
\end{center}
\end{figure}

After degaussing the magnetic shield system, and tuning and locking
the laser, the optical rotation as a function of field applied by the
internal coil system was measured; the results are displayed in
Fig.~\ref{fig:spectrum}.  The amplitude of the optical rotation was
$\pm 9$~mrad.  The width of the NMOR feature was characterized by
fitting the data to the derivative of a Lorentzian.  The
valley-to-peak field separation was thereby determined to be 400~pT.
The results are similar to previous work by
others~\cite{bib:budker1998}.

The definition of zero optical rotation was not carefully controlled
in these measurements; it was adjusted periodically to ensure the
signal from the balanced photoreceiver was near zero.  The vertical
scale of Fig.~\ref{fig:spectrum} therefore contains a small arbitrary
offset.  The horizontal positioning of the zero-field point in the
graph refers to zero current applied by the internal solenoidal coil.
This is displaced by 15~pT from the steepest point of the NMOR data
because of the remanent magnetization of the magnetic shield system,
which we did our best to remove via degaussing.

The dispersive feature in Fig.~\ref{fig:spectrum} could be narrowed
somewhat by applying constant fields with the internal saddle coils
(described in Sec.~\ref{sec:coils}) as the $z$-coil was swept through
zero field.  The narrowest spectrum was found by applying an internal
transverse field of 130~pT,
implying that after degaussing, a residual transverse field of order
this magnitude remained.  For the measurements presented in this
paper, the transverse coils were switched off and disconnected from
their power supply in order to avoid additional sources of noise.  The
method of calibrating optical rotation to the current in the internal
$z$-coil makes our method insensitive to static transverse fields,
excepting that the magnetometer sensitivity is reduced slightly in a
non-zero transverse field.  The calibration and method is described in
Section~\ref{sec:shielding}.

We found we were able to better degauss our shields by determining the
remanent longitudinal and transverse fields by requiring graphs like
Fig.~\ref{fig:spectrum} cross zero as close as possible to zero
applied current, and be as narrow as possible for zero applied
transverse field.  The status of the shields for this measurement
represents our best efforts to use the degaussing system in this
fashion.

The bandwidth of the magnetometer was measured by applying a
sinusoidal magnetic field using the internal $z$-coil, of amplitude
0.5~pT, and measuring the amplitude of optical rotation, which varied
sinusoidally at the same frequency.  The amplitude of the optical
rotation signal was found to drop to $1/\sqrt{2}$ its value at low
frequencies for an applied frequency of 2.7~Hz, indicating the -3~dB
point of the magnetometer.

\begin{figure}
\begin{center}
\includegraphics[width=\textwidth]{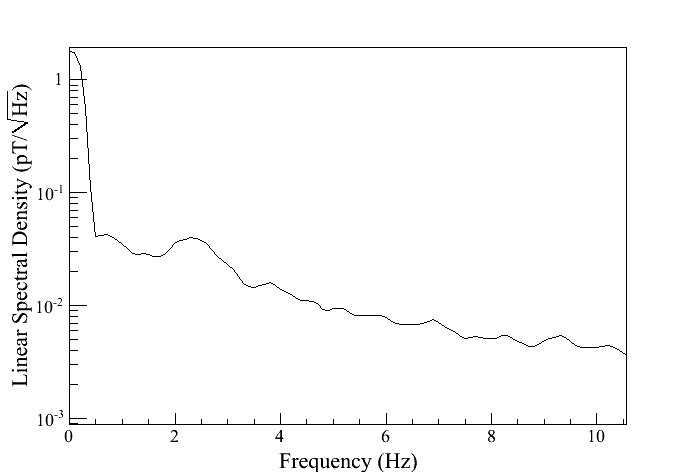}
\caption{Linear spectral density of noise in the magnetometer when all
  coils systems are disconnected.  The window function HFT116D was
  used, with the overlap recommended in
  Ref.~\cite{bib:lisapathfinder}.\label{fig:lsd}}
\end{center}
\end{figure}

Noise in the magnetometer was characterized by measuring the linear
spectral density, with the internal coils switched off and
disconnected.  A graph of the linear spectral density is shown in
Fig.~\ref{fig:lsd}.  The noise was found to be 30~fT/$\sqrt{\rm Hz}$
at a frequency of 1~Hz.  A roll-off in noise is seen above 2.5~Hz, as
expected, since the sensitivity of the magnetometer is reduced above
these frequencies.  A peak is seen for frequencies $<0.4$~Hz due to
uncorrected DC offsets.

The noise of the magnetometer was also estimated by observing the
optical rotation for very small fields applied in a square-wave using
the internal $z$-coil, with the bandwidth of the balanced polarimeter
limited to 1~Hz using a Stanford Research Systems low-noise
preamplifier (Model SR560)~\cite{bib:srs}.  The smallest change in the
DC field that could be observed reliably was of order 40 fT under
these conditions, consistent with the determination based on the
spectral density.

\section{Axial shielding factor measurement using very small applied field\label{sec:shielding}}

An external axial magnetic field was applied to the shield using two
circular coils of diameter 30~cm and spaced by 45~cm.  Each coil was
formed from a single loop of 24~AWG wire.  The coils were spaced
symmetrically from the endcaps of the magnetic shield.  The axial
field generated at the center of the coils in the absence of the
shield was measured to be 1.45~$\mu$T/A with a fluxgate magnetometer,
in agreement with expectation.

\begin{figure}
\begin{center}
\includegraphics[width=\textwidth]{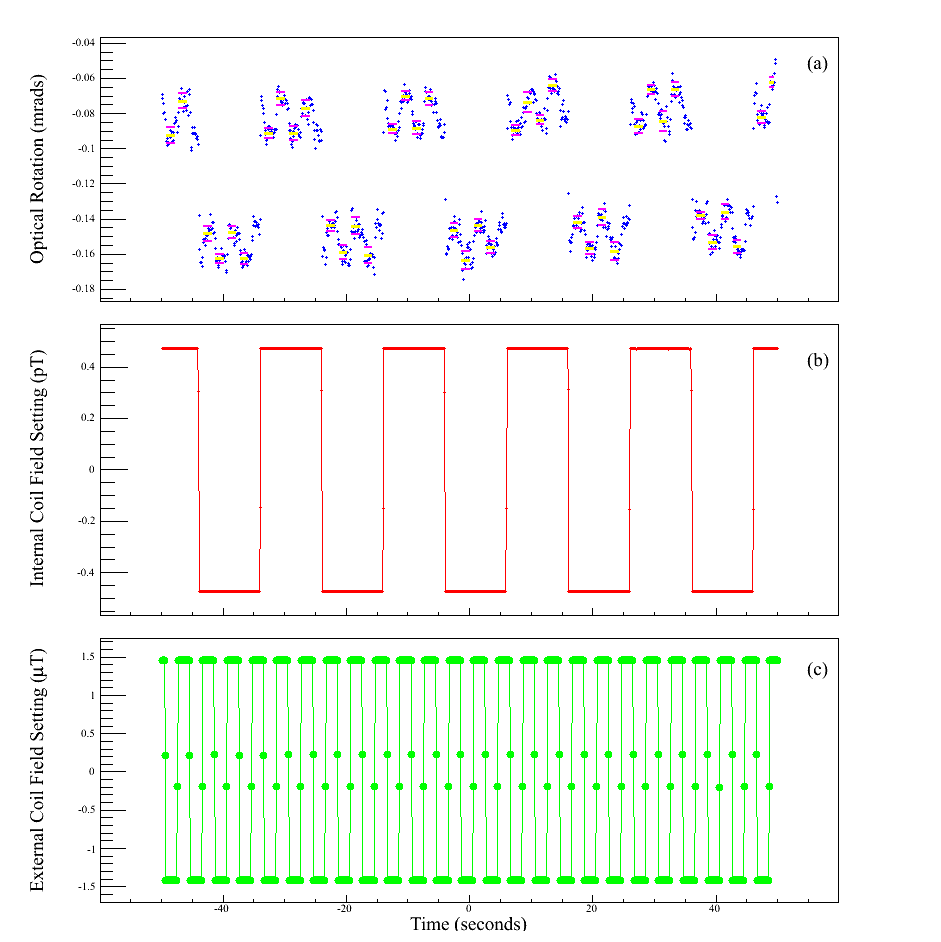}
\caption{(a) Optical rotation, (b) field generated by the internal $z$
  coil, and (c) field generated by the external coil, as a function of
  time.  The optical rotation is seen to depend on both applied
  fields.  Since the field generated by the internal coil is known,
  the shielding factor for the external field may be
  determined.\label{fig:shielding}}
\end{center}
\end{figure}

In order to measure the axial shielding factor, the internal $z$-coil
and the external coil were operated by different square-wave function
generators and the optical rotation was observed as a function of
time.  The internal coil was set to provide a field of $\pm 480$~fT
with a period of 20~s.  The external coil was set to provide a current
of $\pm 1$~A (resulting in a central field $B_c=\pm 1.45\mu$T in the
absence of the shield) with a period of 4~s.  Fig.~\ref{fig:shielding}
displays the optical rotation measured as a function of time during
this measurement.  Since the internal $z$-coil calibration is known,
this data enables a determination of the axial field produced
internally by the external coils, under the assumption that optical
rotation is linear in the axial magnetic field (an assumption that was
confirmed by calibrating using different internal $z$-coil currents).

The optical rotation for each setting of the internal and external
field in Fig.~\ref{fig:shielding} was averaged.  Data within 0.5~s of
a transition in the external field were excluded.  Data within the 2~s
stable period of the external field, but containing a transition in
the internal field, were excluded entirely.  Typically fourteen
optical rotation data points were averaged in each stable period.  The
result of the averaging was sensitive to changes in these time cutoffs
at the $< 10\%$ level.  The standard deviation for each time period
was also determined, and found to be typically 0.005~mrad.  In
determining the measured fields, successive pairs of measurements were
considered in order to eliminate the effect of long-term drifts.  The
field generated internally by the external coils was thereby
determined to be $\pm 110$~fT, or a span of 220~fT.

\subsection{Errors in the Measurement}

The statistical uncertainty in the measurement comparing successive
time-periods in the data is estimated from the standard deviation
divided by the square-root of the number of measurements in each
period to be 10\%.  Fifty time periods in the data were averaged,
representing twenty-five statistically independent pairs, and the
statistical error on the change in internal field is 2\% if all time
periods are used.

Measurements of optical rotation were occasionally seen to drift over
the typical 100~s measurement cycle.  In some runs, the drift over
100~s was found to be as large as 0.04~mrad.  The source of the drift
is unclear at this time, whether a true magnetic change in the shield
system (corresponding to a change of $\sim$0.6~pT over the same time
period) or a drift in the magnetometer itself.  From data taken with
all coils disconnected, when drifts occur, the Allan standard
deviation is seen to rise for timescales above 1~s.  This implies that
drifts start to dominate the error for timescales longer than seconds.
Taking the 2~s averaging time that we used, with a worst-case drift of
0.04~mrad/100~s the error due to drift within the averaging time would
be 5\%.

Based on this analysis, considering statistics, drifts, and dependence
on time cuts used in the data, the error in the 220~fT internally
induced field change from the external coils is $10\%$ or 20~fT.

We defined the axial shielding factor as the reduction of the central
$B_c=\pm 1.45~\mu$T produced by the external coils in the absence of
the magnetic shield system to the central field produced at the same
position in the presence of the shield.  This would agree well with an
average over the vapour cell volume (to better than 1\%), since the
cell dimensions are smaller than the nearest magnetic elements in the
system.  The shielding factor is then determined from the ratio
($2.9~\mu$T)/(220~fT)=$1.3\times 10^7$.

%


\subsection{Additional Systematic Studies}

The linearity of optical rotation with axial magnetic field was tested
by varying the amplitude of the internal $z$-coil field between 200
and 1000~fT; the optical rotation was found to be linear to within the
precision of the measurement of the internal field.  The optical
rotation measured by the balanced polarimeter was calibrated to a
physical rotation of a half-wave plate placed in the path of the laser
and found to be in agreement with expectation.

The field generated by the external coil in our experiment is not
uniform; a larger external coil system was not possible in our setup,
because of physical constraints and nearby magnetic elements.  A
concern in the measurement was that the shield experiences an
effectively larger field than the central field of the external coil
system would indicate, because of the closeness of the coils to the
outermost shield.  To test this, the spacing between the coils was
increased, thereby increasing the distance of the coils from the
outermost shield.

\begin{table}[htdp]
\begin{center}
\begin{tabular}{ccccc}
\toprule
Configuration & Spacing (cm) & Current (A) & $B_c$ ($\mu$T) & SF ($\times 10^6$)  \\
\noalign{\smallskip}
\midrule
\noalign{\smallskip}
1& 45 & $\pm 1.0$ & $\pm 1.45$ & $13.1\pm 1.3$ \\
\noalign{\smallskip}
\midrule
\noalign{\smallskip}
2 & 50& $\pm 1.0$ & $\pm 1.15$ & $13.8\pm 1.8$ \\
\noalign{\smallskip}
\midrule
\noalign{\smallskip}
3 & 60& $\pm 1.0$ & $\pm 0.75$ & $14.7\pm 3.2$ \\
\noalign{\smallskip}
\midrule
\noalign{\smallskip}
4 & 60& $\pm 2.0$  & $\pm 1.50$ & $13.3\pm 1.3$ \\
\noalign{\smallskip}
\bottomrule
\end{tabular}
\end{center}
\caption{Shielding factor (SF) for the various configurations of
  $z$-coil spacing and current.  The nominal central field $B_c$ in
  the absence of the shield was determined numerically and verified by
  fluxgate measurement to within 5\%.  The uncertainty of 20~fT on the
  measurement of the internal field via NMOR dominates the error in
  SF.\label{tab:configs}}
\end{table}

Table~\ref{tab:configs} displays the resultant shield factor
determined for various trials where the spacing of the external coils
and the current flowing through the external coils was changed.  For
changes to the geometry of the coil admitted by the physical space
limitations of our experiment (corresponding to coil spacings ranging
from 45 to 60 cm), the shielding factor was found to vary by 8\%,
within the stated errors on the measurements.  Since the measurements
are all in agreement, we retain the result for the shielding factor
$(1.3\pm 0.1)\times 10^7$.

\subsection{Earth's field, other transverse fields, and limitations}

The magnetic shield and coil system were always subject externally to
Earth's magnetic field, which is considerably stronger than the $\pm
1.45\mu$T field supplied by the external coil system.  The Earth's
field was dominantly transverse to the axis of the laser beam and
shield.  In a model where the magnetic material in the shield is
linear, the presence of a transverse field of this sort does not
affect the axial shielding factor measured using our technique.
However, if the material is not linear, the results could depend on
the saturation of the material.  Our magnetic shielding factor
measurements therefore possess this caveat.

Finally, it is possible that the application of an external field by
our coil system could generate internal fields that are not entirely
longitudinal when averaged over the volume of the vapour cell.  This
could occur due to any lack of cylindrical symmetry in the system,
resulting from e.g.~misalignment and imperfections in the magnetic
shielding.  Furthermore the magnetometer, and our calibration scheme,
are sensitive to transverse fields.  We measured this sensitivity by
applying similar magnitude ($\pm 250$~fT) transverse fields using the
internal saddle coils and found the optical rotation to be, at
largest, a factor of two smaller for the same field applied axially,
and dependent on the direction of the transverse field.  Thus, while
the sensitivity was smaller, it was not significantly smaller.  It is
therefore an assumption of the procedure that only the axial field
changed when exciting the external coil.

\section{Conclusion}

We have employed an NMOR-based magnetometer to make measurements of
small axial magnetic fields generated within a cylindrical
magnetically shielded volume by a pair of external circular coils.
The shielding factor for an axially-directed 1.45~$\mu$T field was
determined to be $(1.3\pm 0.1)\times 10^7$, in the background of
Earth's field.  The result was consistent for even lower applied
fields of 0.75~$\mu$T.  The results reproduce the design expectations
of the magnetic shield, which were based on finite element analysis.
The results therefore provide a useful benchmark for future magnetic
shield design.

The results highlight the applicability of NMOR-based magnetometers
for small fields inside magnetically shielded volumes, such as those
that will be experienced in future neutron electric dipole moment
experiments.  In the future we plan to study the stability of
NMOR-based magnetometers and the stability of uniform magnetic fields
generated by coils internal to the measurement volume, similar to the
situation in future EDM experiments.

\section{Acknowledgements}

We would like to thank D.~Budker and B.~Patton for their advice on the
development of the magnetometer, and for the loan of their vapour
cell.  This work was supported by the Canada Foundation for
Innovation, the Natural Sciences and Engineering Research Council
Canada, and by the Canada Research Chairs program.


\section*{References}


\end{document}